\documentclass{emulateapj}
\usepackage{natbib}
\usepackage{graphicx}
\usepackage{amssymb}
\shorttitle{Close Pairs as Proxies for Mergers}
\shortauthors{Wetzel et al.}

\begin{document}
\title{Close Pairs as Proxies for Galaxy Cluster Mergers}
\author{Andrew R. Wetzel\altaffilmark{1}, A.E. Schulz\altaffilmark{1}, 
Daniel E. Holz\altaffilmark{2,3}, and Michael S. Warren\altaffilmark{2}}
\altaffiltext{1}{Department of Astronomy, University of California, 
Berkeley, CA 94720}
\altaffiltext{2}{Theoretical Division, Los Alamos National Laboratory, 
Los Alamos, NM 87545}
\altaffiltext{3}{Department of Astronomy \& Astrophysics, 
University of Chicago, Chicago, IL 60637}

\begin{abstract}
Galaxy cluster merger statistics are an important component in understanding the
formation of large-scale structure.
Cluster mergers are also potential sources of systematic error in the mass
calibration of upcoming cluster surveys.
Unfortunately, it is difficult to study merger properties and evolution directly
because the identification of cluster mergers in observations is problematic.
We use large N-body simulations to study the statistical properties of massive
halo mergers, specifically investigating the utility of close halo pairs as
proxies for mergers. 
We examine the relationship between pairs and mergers for a wide range of merger
timescales, halo masses, and redshifts ($0<z<1$).
We also quantify the utility of pairs in measuring merger bias.
While pairs at very small separations will reliably merge, these constitute a 
small fraction of the total merger population. 
Thus, pairs do not provide a reliable direct proxy to the total merger 
population.  
We do find an intriguing universality in the relation between close pairs and 
mergers, which in principle could allow for an estimate of the statistical 
merger rate from the pair fraction within a scaled separation, but including 
the effects of redshift space distortions strongly degrades this relation.
We find similar behavior for galaxy-mass halos, making our results applicable 
to field galaxy mergers at high redshift.
We investigate how the halo merger rate can be statistically described by the
halo mass function via the merger kernel (coagulation), finding an interesting
environmental dependence of merging: halos within the mass resolution of our
simulations merge less efficiently in overdense environments.
Specifically, halo pairs with separations less than a few $h^{-1}\,\mbox{Mpc}$ 
are more likely to merge in underdense environments; at larger separations, 
pairs are more likely to merge in overdense environments.
\end{abstract}

\keywords{cosmology:theory -- methods:numerical -- dark matter: merging
histories -- galaxies: clusters}

\section{Introduction}
Galaxy clusters are of great interest in cosmology as they are the largest
and most recently formed structures in the cosmological hierarchy.
The clustering and number density evolution of clusters are sensitive to both
the growth function and the expansion history of the universe.  
Clusters contain a representative sample of baryons and dark matter and are thus
also fascinating laboratories in which to study the influence of baryonic
physics on the formation of large-scale structure, substructure, and its effect
on the gravitational potential and dark matter halo shape.
Statistical measures such as the mass function, the rate of structure growth, 
and the clustering of large structures, are fundamental predictions of 
cosmological models.
Cluster observations are therefore expected to provide some of the most 
important constraints on fundamental cosmology and astrophysics 
(see e.g.~\citet{Bor06} for a recent review).

Cluster formation histories are frequently punctuated by large jumps in mass
from major mergers (e.g.~\citet{CohWhi05}).
These mergers are one of the primary mechanisms for the buildup of mass in
clusters and superclusters.
In the standard paradigm \citep{Kai84, Efs88, ColKai89, MoWhi96, SheTor99},
observable properties such as the degree of spatial clustering depend
\textit{only} on the cluster mass.
However, recent theoretical studies indicate that many cluster observables, 
such as spatial clustering, concentration, galaxy velocity dispersion, gas 
distribution and its attendant observables such as X-ray emissions and SZ 
decrement, depend on the cluster's formation time, mass accretion history, 
large scale environment (collectively referred to as ``assembly bias'') 
\citep{Wec02, Zha03a, SheTor04, GaoSprWhi05, Wec06, CroGaoWhi06, Har06, Wet07, 
GaoWhi07, WanMoJin07, JinSutMo07, Hahn07, Mac07, San07}. In addtion, there is a
dependence on recent merger history (``merger bias'')
\citep{ScaTha03, RowThoKay04, FurKam05, Wet07, Poo07, Jel07, Har07}.
In addition, recent studies have claimed observational detection of assembly 
bias, though with mixed results \citep{Berl06, YaMoBo06}.
Theory can predict cluster properties as a function of their mass, which is
dominated by dark matter and thus cannot be directly measured.
Since all methods of observing clusters are sensitive to the effects of
assembly/merger bias, it is necessary to develop a more detailed understanding 
of the mechanisms of structure formation.
Specifying a correlation function and mass function, and the evolution of these
quantities, may not be sufficient to connect theory to observation. 
Understanding cluster merger properties is therefore crucial to utilize these 
objects as probes of cosmology.

We focus primarily on galaxy cluster mergers.
\citet{Wet07} used a large simulation volume to probe high-mass 
halos with good statistics.
They found that halos of mass $M>5\times10^{13}\,h^{-1}M_{\odot}$ that have 
undergone a recent (within 1 Gyr or less) major merger or large mass gain
exhibit an enhancement in their spatial clustering of up to $\sim10$--$20\%$ on 
scales of $5$--$25\,h^{-1}$Mpc compared with the entire halo population at the 
same mass.
They noted that this merger bias persists for the redshift range $0<z<1$, and 
that the bias increases with larger merger mass ratios or shorter merger
timescales.
If this merger bias remains unaccounted for, then mass-observable relations 
that connect theory with observation calibrated on the basis of clustering may 
be suspect.

Moreover, the gas properties of a recently merged halo cluster can be quite
different from the general population.
This can be mitigated by selecting ``relaxed'' halos (presumably those that 
have not had recent major mergers) when calibrating observables.
However, the scaling relations from this selected sample may be biased with
respect to that from the overall cluster population.
In addition, if a merger accidentally entered the ``clean'' sample 
(e.g.~because it occurred along the line of sight), it could substantially bias
the calibration, and hence the ensuing mass determination.
It is thus important to predict the fraction of halos which have had recent 
merger activity.
Observing cluster mergers as a function of redshift, studying the merger bias,
and correlating mergers with other tracers of density and environment will also
shed light on the nature of structure formation. 

A number of authors (e.g.~\cite{Pat97, Pat00, Pat02, Dep05, Inf02, Lin04, Con06,
Bel06, Mas06, Dep07}) have studied the statistics of galaxy merging using close 
pairs as a proxy for mergers. 
In this work we extend those analyses to the cluster scale.
It is important to quantify the conditions for which the pair proxy assumption 
is valid.
\cite{Ber06} used simulations to examine the utility of galaxy pairs 
in measuring the redshift evolution of halo merger rates out to $z \approx 3$.
They tracked the formation and evolution ``host'' halos of dark matter, as 
well as self-bound density peaks within host halos, i.e. ``subhalos,'' a term
which includes both satellite objects and the central host itself.
Using the assumption that galaxies are found at the centers of subhalos, 
they found that galaxy pair counts can probe the rates at which satellite 
galaxies merge with the central galaxy, i.e. \textit{subhalo} mergers.
These pairs can also constrain the galaxy Halo Occupation Distribution, i.e. 
the statistical number of galaxies found within a host halo of a given mass. 
However, galaxy pairs cannot be used to measure the merger rates of separate
massive \textit{host} halos.

The goal of this paper is to investigate whether close spatial pairs of halo 
objects (galaxy clusters) can be used to observationally study the properties 
of mergers.
Identifying a cluster merger in an observation is not straightforward because it
relies principally on the object appearing morphologically ``disturbed''
\citep{RowThoKay04}.
Furthermore, it is difficult to estimate the completeness and the contamination
of such an observed merger population.  
If close pairs prove to be a sufficient proxy for mergers, they would provide a 
clean tool to study merger statistics.
We concentrate primarily on massive halos ($M > 5\times10^{13}\,h^{-1}M_\odot$),
corresponding to galaxy groups and clusters, in contrast to previous work 
which has focused on pairs as a proxy for galaxy mergers.
However, our results also apply to any class of object that singly occupies a
halo, particularly field galaxies at high redshift.
 
In \S\ref{sec:sim} we describe the simulations and halo catalogs used in our
study.
In \S\ref{sec:close} we present statistics for merger rates as a function of
halo pair masses, pair separations, and merger time interval to quantify the
extent to which close pairs can be used as proxies for cluster mergers in the
best-case scenario where halo spatial positions are known accurately.
In \S\ref{sec:redspace} we consider observational complexities such as 
scatter in the halo mass, redshift space distortions, and redshift space errors.
In \S\ref{sec:kernel} we present a mass-dependent fit to the merger 
kernel and examine how large-scale density impacts the efficiency of pair
merging.  
In \S\ref{sec:xipairs} we discuss the prospects for studying merger bias using
close cluster pairs.
We conclude in \S\ref{sec:conclusions}.

\section{Simulations} \label{sec:sim}

Our study is conducted using two high-resolution N-body simulations of cold dark
matter in a flat $\Lambda$CDM cosmology with parameter values $\Omega_M=0.3$,
$\Omega_B=0.046$, $h=0.7$, $n=1$ and $\sigma_8=0.9$.  
Our simulations employ the HOT code \citep{HOT} in a ($1.1\,h^{-1}\mbox{Gpc})^3$
and a ($2.2\,h^{-1}\mbox{Gpc})^3$ volume with periodic boundary conditions, 
using a Plummer softening length of $35\,h^{-1}\,\mbox{kpc}$.
Gaussian initial conditions were randomly generated for the $1024^3$ particles
of mass $10^{11}\,h^{-1}M_\odot$ (smaller simulation) and 
$8 \times 10^{11}\,h^{-1}M_{\odot}$ (larger simulation) at an initial redshift
of $z=34$.  
Simulation outputs were stored in intervals of $1$~Gyr between redshifts
$z\approx 1$ and $z=0$, with the last interval of each simulation being shorter:
$0.6$~Gyr (smaller simulation) and $0.5$~Gyr (larger simulation).
All time intervals cited below represent the total time elapsed between two 
simulation outputs.
Thus a merger timescale of $\Delta t = 1$~Gyr indicates that two separate halos
have merged within 1 Gyr, an upper limit to the actual time to merger.

We generate a halo catalog for each output using the Friends-of-Friends (FoF)
algorithm \citep{DEFW} with a linking length of $b=0.15$ of the mean
interparticle spacing.
These groups correspond to a density threshold of $\sim3/(2\pi b^3)$ and enclose
primarily virialized material.
In this work we will quote FoF masses, which are about 20\% smaller than
``virial'' masses (corresponding roughly to FoF masses with $b=0.2$, which are
more commonly found in the literature; see \citet{Whi01} for more details).
Since we are examining mergers, we use a smaller linking length to decrease
contamination that arises from close neighboring halos being bridged by a narrow
string of particles (although using a larger linking length of $b=0.2$
changes our results by only a few percent, see Fig.~\ref{fig:fmergemass}). 
The halo catalogs of the smaller simulation include all halos of mass greater
than $M \approx 5 \times 10^{12}\,h^{-1}M_{\odot}$ ($>50$ particles), though in
our study we are concerned primarily with halos of mass  
$M>5\times 10^{13}\,h^{-1}M_{\odot}$, of which there are around
$75,000\,h^{-3}\mbox{Gpc}^{-3}$ at $z=0$.  
Our larger ($2.2\,h^{-1}\mbox{Gpc})^3$ but less resolved simulation catalogs 
include all halos of mass greater than 
$M \approx 4 \times 10^{13}\,h^{-1}M_{\odot}$ ($>50$ particles). 
We do not consider substructure within host halos.

Merger trees were constructed from the set of halo catalogs by specifying a
parent-child relationship, where a ``parent'' is any halo that contributed mass
to a halo at a later time, i.e.~a ``child.''
We define a parent contributing more than half of its mass as a ``progenitor.''
Under this restriction, a progenitor can never have more than one child. 
We define a merger as a child having more than one progenitor.
In cases where we select on a child that had more than two progenitors, we apply
the two body approximation by considering only the two progenitors that
contributed the most mass to the child.
Using instead the two most massive progenitors, as opposed to the two most
mass-contributing progenitors, changes the progenitor identification in less
than 1\% of all mergers that we consider.
In addition to this progenitor merger tree, a list of all contributing parents
(not just progenitors) is stored for each child.
This flexible storage of parent data allows us to study two body mergers,
mergers with more than two progenitors, or mass accretion including all parents.
However, when considering short timescales ($\lesssim 1$~Gyr), the two body
criterion is a good approximation (see \citet{Wet07} for more details).

All errors cited are $1\sigma$ errors derived from dividing the simulations into
8 octants and computing the dispersion of the quantity of interest within each
octant.
Since we probe scales much smaller than the octants, we treat them as
uncorrelated volumes.

\section{Close Pairs as Predictors of Mergers} \label{sec:close}

\begin{figure}
\begin{center}
\resizebox{3.1in}{!}{\includegraphics{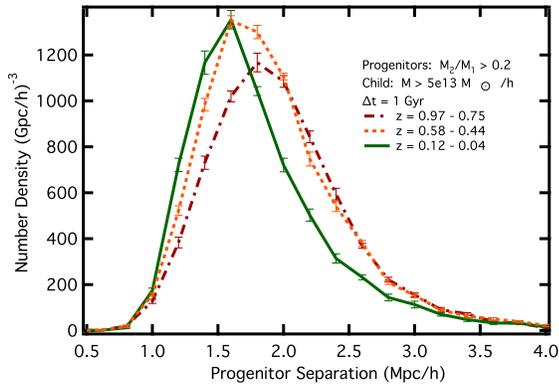}}
\end{center}
\caption{Number density of progenitor pairs as a function of (comoving) binned
separation.  
We consider children with masses above $5\times 10^{13}\,h^{-1}M_{\odot}$ and
restrict ourselves to cases where the two most mass-contributing progenitors
have total mass ratios $M_2/M_1 > 0.2$ (major mergers).  
A fixed merger time interval of $\Delta t = 1$~Gyr is used for child halos at
$z=0.04$ (solid), $z=0.44$ (dotted), $z=0.75$ (dot-dash).}
\label{fig:parsepz}
\end{figure}

To examine whether close pairs of galaxy clusters form reliable predictors
of mergers, we first extend the work of \cite{Wet07} on halo mergers, using
the same simulation and halo catalogs (described in \S\ref{sec:sim}).  
We identify child halos of mass $M>5 \times 10^{13}\,h^{-1}M_{\odot}$ that
are products of recent mergers with progenitor masses $M_1$ and $M_2$, 
where $M_2/M_1 > 0.2$ (major mergers).
We explore the distribution of progenitor separations in Fig.~\ref{fig:parsepz},
which shows the number density of progenitors as a function of binned progenitor
separation.
For time intervals of $1$~Gyr, there is a characteristic comoving progenitor
separation ($\sim1.5\,h^{-1}\mbox{Mpc}$) that does not evolve significantly with
redshift.  
Without the influence of gravitational attraction, halos with
typical velocities of $\sim\!1000\,\mbox{km}/\mbox{s}$ will travel $1$~Mpc 
within $1$~Gyr.
Furthermore, the number density of mergers evolves only weakly with redshift.
These factors suggest that close pairs might be a reasonable proxy for mergers.

These results represent a post-diction, where we know in advance which
halos will merge.  
We now investigate whether close pairs of objects can reliably be used to
\textit{predict} mergers.
The following subsections investigate how frequently pairs will merge as a
function of pair mass, (comoving) separation, redshift, and merger time
interval.
We consider a given pair to have merged if both halos are progenitors of the
same child at a later timestep.
Since observations of galaxy clusters are generally limited by a threshold
luminosity, we identify pairs of halos whose individual halo masses are above a
given mass cut.
We first examine the pair-merger hypothesis (that close pairs are merger 
proxies) for the best-case scenario in which halo positions and masses are 
known with complete accuracy (as in our simulations).
This will provide a firm upper limit to the utility of pairs in predicting 
mergers.
Then in \S\ref{sec:redspace} we consider observational complexities such as 
scatter in the halo mass, redshift space distortions, and redshift space errors,
which significantly degrade the signal.

\subsection{Pair Mergers at $z=0$}

\begin{figure}
\begin{center}
\resizebox{3.1in}{!}{\includegraphics{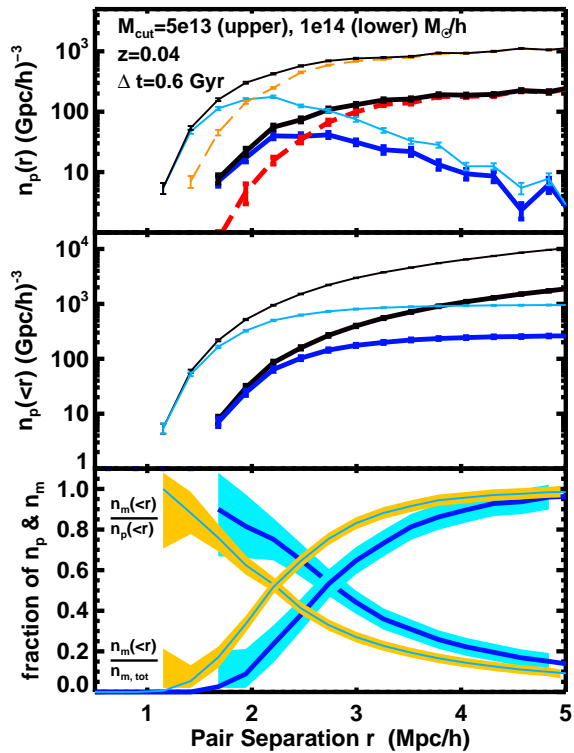}}
\end{center}
\caption{{\bf Top:} The number density of pairs at $z=0.04$ as a function
of binned separation, $n_p(r)$ (black solid), the subset of these that
merged in $\Delta t=0.6$~Gyr, $n_m(r)$ (blue solid), the remainder that did
not merge (red dashed).
The upper curves (thin) are for pairs of mass 
$M > 5 \times 10^{13}\,h^{-1}M_{\odot}$, and 
the lower (thick) are for more massive pairs, 
$M > 10^{14}\,h^{-1}M_{\odot}$.   
The intersection with the dashed curve shows the separation beyond which the
majority of pairs do not merge, highlighting the limited range of separations
for which merging is likely.
{\bf Center:} The same, but integrated for all separations smaller than the
abscissa. 
{\bf Bottom:} The fraction of pairs separated by less than $r$ that merged,
${n_m(<r)}/{n_p(<r)}$ (falling curves), indicating the efficiency of
pairs as merger candidates, and the ratio of identified mergers to all
mergers above the mass cut, ${n_m(<r)}/{n_{m,tot}}$ (rising curves),
indicating the completeness of the candidate sample.
Error bars and shaded regions indicate the $1\sigma$ error of cosmic
variance from dividing the simulation into octants.}
\label{fig:mergerstat_zis0}
\end{figure}

We select halos at $z=0.04$ and consider mergers to $z=0$ ($\Delta t =
0.6$~Gyr).
Figure~\ref{fig:mergerstat_zis0} (top) shows the number density of pairs as
a function of pair separation, $n_p(r)$, along with the number densities of
pairs that merged (solid), $n_m(r)$, and did not merge (dashed)
within the time interval.
The number density of pairs terminates at small separations because
of halo exclusion, i.e.~massive halos have finite radii.
The upper set of curves are halos of mass $M>5\times 10^{13}\,h^{-1}M_{\odot}$
and the lower have $M>10^{14}\,h^{-1}M_{\odot}$; the results are qualitatively
similar.
There is a limited range of separations larger than the halo exclusion limit in
which the majority of pairs merge.
Also plotted (center) are the same number densities, but as an integrated
function of separation, demonstrating the relationship for pairs
\textit{within} the given separation $r$.  
In $0.6$~Gyr at $z=0$, massive halo pairs will only merge with certainty
for separations $\lesssim 2\,h^{-1}$Mpc.

Figure~\ref{fig:mergerstat_zis0} (bottom) also shows the fraction of pairs that
merge \textit{within} the given separation, ${n_m(<r)}/{n_p(<r)}$ (falling
curves).
This fraction is the  ``efficiency'' of the close-pair method in identifying
merger candidates, since it shows the likelihood that a pair within the given 
separation will merge. 
It can also be thought of as a measure of contamination of the candidate sample,
because its difference from $1$ identifies the fraction pairs that do not merge.
The rising curve shows the fraction of mergers over the \textit{total} number
density of mergers, ${n_m(<r)}/{n_{m,tot}}$.  
This can be interpreted as the merger ``completeness,'' as it shows the fraction
of the total number of mergers found by identifying pairs within the given 
separation.
Figure~\ref{fig:mergerstat_zis0} confirms the intuitive result that because of 
greater accelerations, larger halos are able to merge from larger separations, 
which holds at all redshifts and time intervals (see Fig.~\ref{fig:fmergemass}).

If close pairs were ideal predictors of mergers, the rising and falling curves 
of Fig.~\ref{fig:mergerstat_zis0} (bottom) would be step functions,
crossing near a fraction ($y$-value) of $1$ and demonstrating a clear
dichotomy between those pairs that merge and those that do not.
However, because of dynamical effects halos do not simply accelerate toward one 
another from rest; they merge with neighbors from a broad distribution of 
separations.
Indeed, the effects of local environment can hinder the merging of close pairs
that would ordinarily occur via non-linear two-body gravitational interaction
(see \S\ref{sec:kernel} for more details).

The intersection of the efficiency and completeness curves represents an easily
identifiable pair separation that compromises between maximizing completeness
while minimizing contamination.
At this intersection the fraction of pairs that merge, ${n_m(<r)}/{n_p(<r)}$, 
equals the fraction of total mergers, ${n_m(<r)}/{n_{m,tot}}$, although the 
relevant populations are different.
If this intersection occurs at a fraction close to $1$, most pairs within the
separation will merge, and those pairs will represent the majority of all
mergers that occur in the time interval.
However, if this fraction is low, pairs within the separation are not a
representative indicator of mergers.
Since the intersection of these two curves occurs at a fraction of $\sim0.5$ for
massive pair mergers within $\Delta t = 0.6$~Gyr at $z=0$, we conclude that
pairs at these redshifts and masses cannot be used to reliably predict mergers.

\subsection{Redshift Dependence of Pair Mergers}

\begin{figure}
\begin{center}
\resizebox{3.1in}{!}{\includegraphics{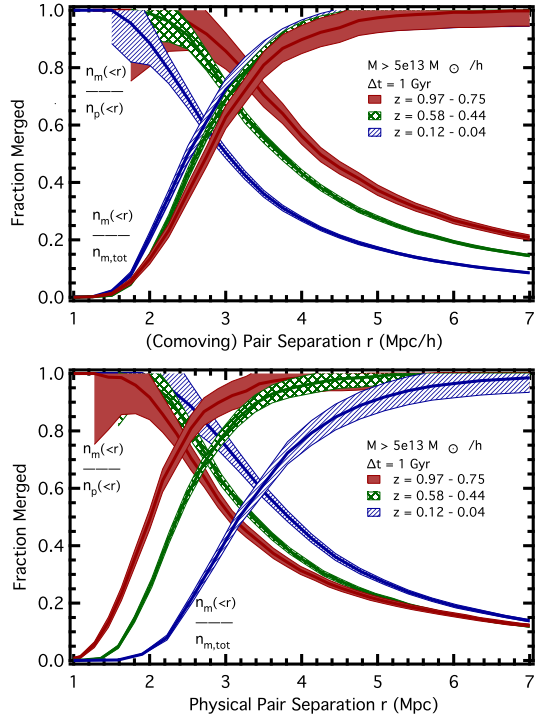}}
\end{center}
\caption{
The fraction of pairs separated by less than $r$ that merged
(falling curves), indicating efficiency, and the ratio of identified
mergers to all mergers above the mass cut within $\Delta t = 1$~Gyr (rising
curves), indicating completeness, 
for various redshifts.
The total number density of such mergers is
$458\,h^{-3}$Gpc$^{-3}$ ($z=0.97-0.75$),
 $950\,h^{-3}$Gpc$^{-3}$ ($z=0.58-0.44$),
 $1464\,h^{-3}$Gpc$^{-3}$ ($z=0.12-0.04$).
Note the approximate invariance of the total merger fraction (rising curve) 
with redshift.
{\bf Bottom:} The same, but as a function of physical separation.
While pairs of a given comoving separation merge more frequently at
high redshift, this trend reverses when considering physical separations.}
\label{fig:fmergez}
\end{figure}

We next examine whether the pair-merger hypothesis fares better at high 
redshift. 
We consider a longer merger timescale of $\Delta t = 1$~Gyr for massive halo
pairs at $z \approx 1, 0.6,$ and $0.1$.
Figure~\ref{fig:fmergez} (top) shows the pair merger fractions as a function of
(comoving) separations.
The pair merger efficiency (falling curves) increases significantly with
redshift, while the merger completeness (rising curve) remains approximately
invariant.
For a given merger timescale, mergers come from pairs of essentially the
same (comoving) separation, regardless of redshift (see Fig.~\ref{fig:parsepz}).
However, since gravitational interactions are governed by physical (not
comoving) separations, Fig.~\ref{fig:fmergez} (bottom) shows the same merger
fractions as a function of physical separation.
The merger efficiency trend is reversed: pairs within a given physical
separation are more likely to merge at low redshift, and mergers come from pairs
of much larger physical separations at low redshift.  
The intersection of the efficiency and completeness curves occurs at a higher
fraction at high redshift than at low redshift (75\% vs 60\%), indicating that,
for a fixed merger timescale, massive halo pairs provide a slightly better proxy
for merger rates at higher redshift.

\subsection{Mass and Linking Length Dependence of Pair Mergers}

\begin{figure}
\begin{center}
\resizebox{3.1in}{!}{\includegraphics{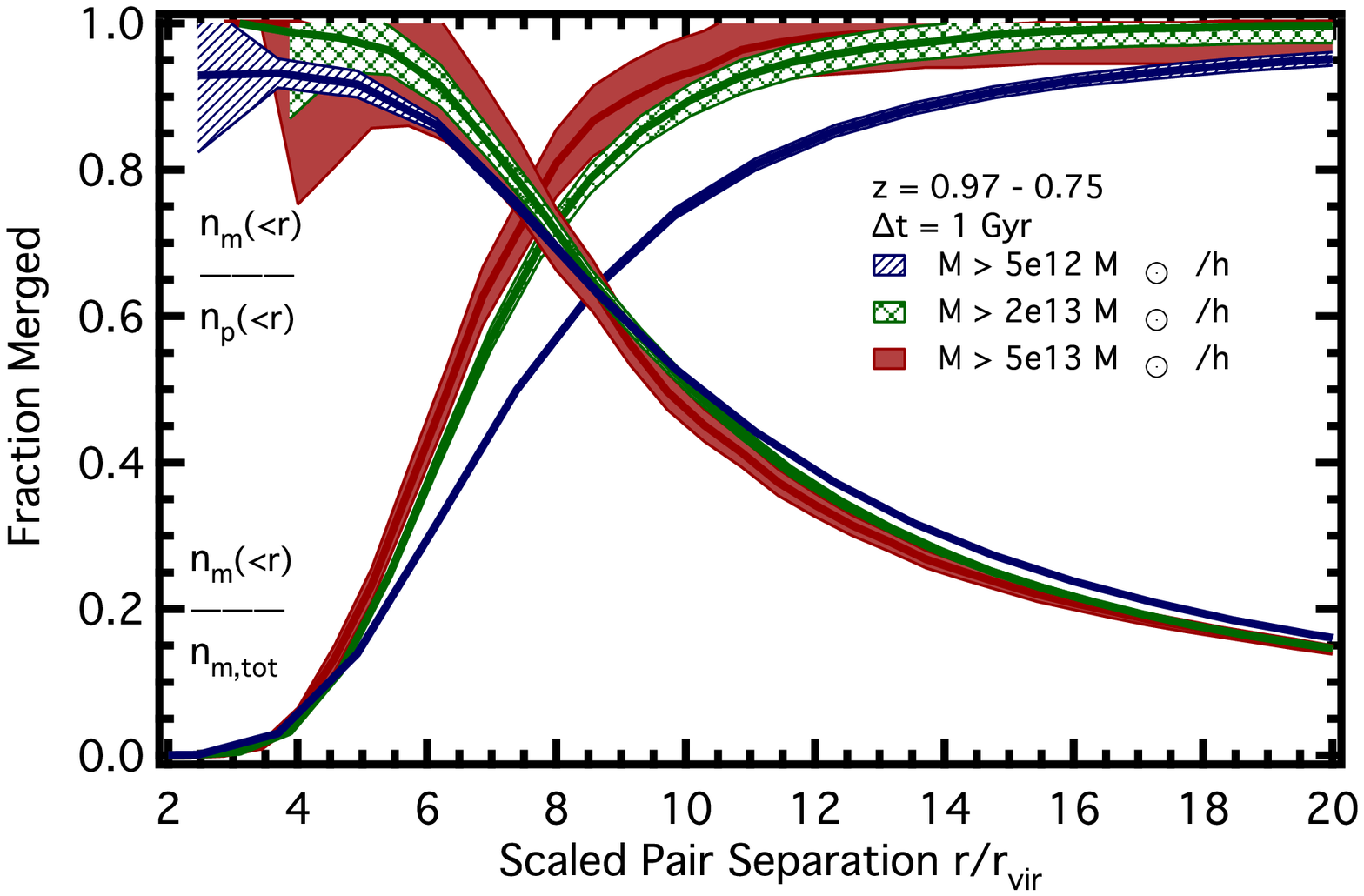}}
\resizebox{3.1in}{!}{\includegraphics{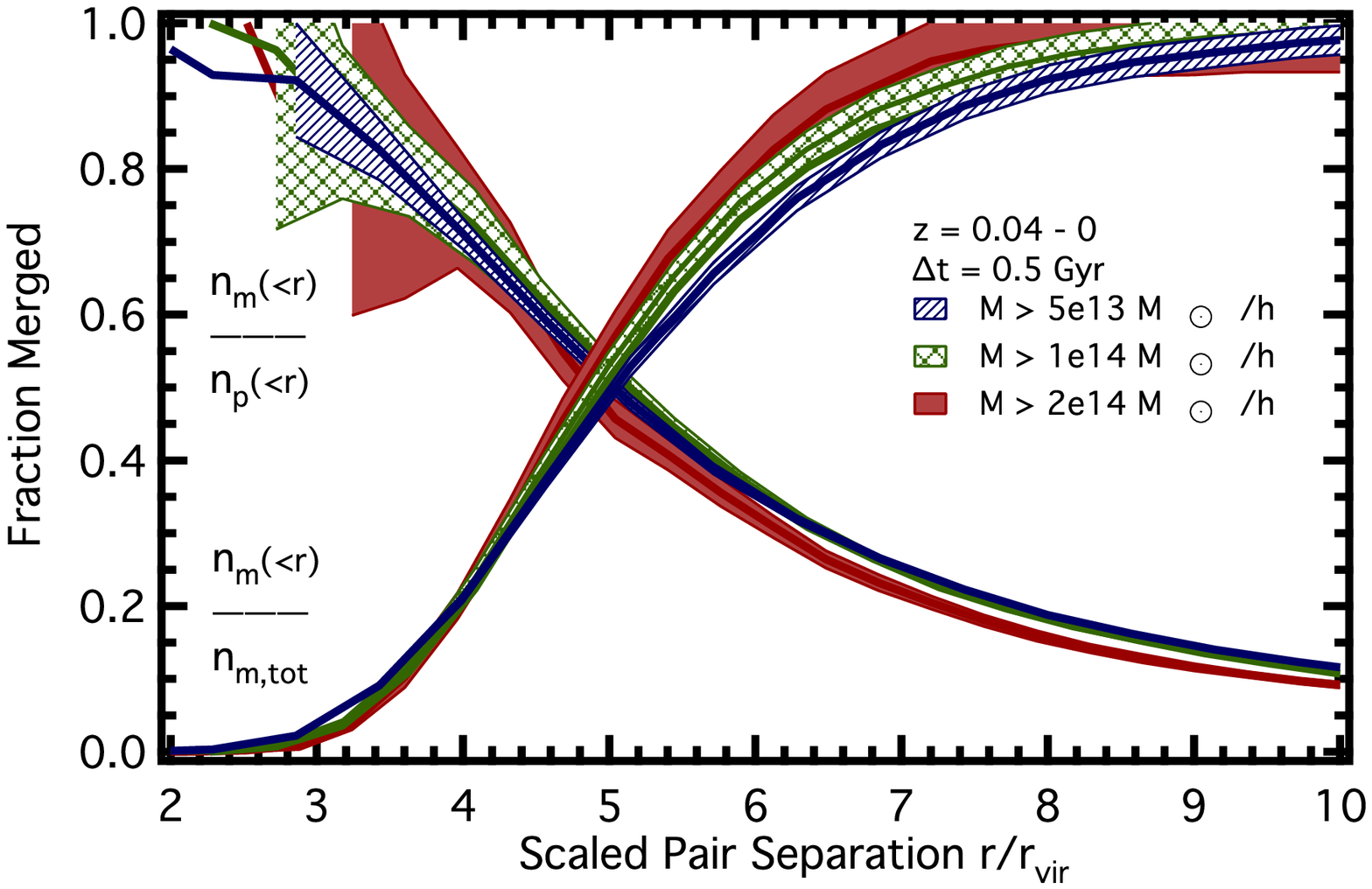}}
\resizebox{3.1in}{!}{\includegraphics{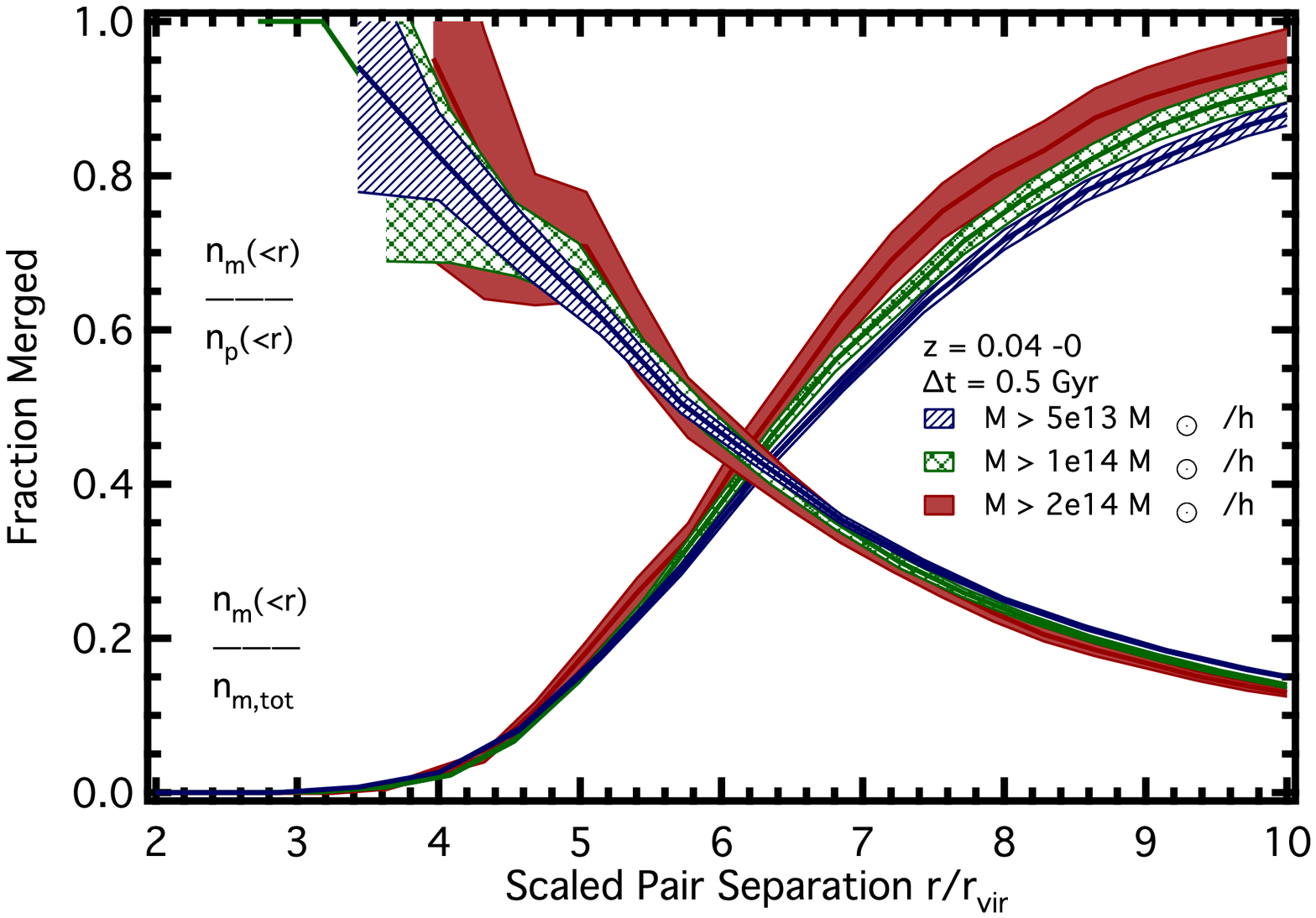}}
\end{center}
\caption{
{\bf Top:} The fraction of pairs separated by less than $r/r_{\mbox{vir}}$ that
merged (falling curves), indicating efficiency, and the ratio of identified
mergers to all mergers above the mass cut within $\Delta t = 1$~Gyr (rising
curves), indicating completeness, at $z \approx 1$ for various mass cuts.
The total number density of such mergers is
$45,750\,h^{-3}$Gpc$^{-3}$ ($M>5\times 10^{12}\,h^{-1}M_{\odot}$),
$3957\,h^{-3}$Gpc$^{-3}$ ($M>2\times 10^{13}\,h^{-1}M_{\odot}$),
$458\,h^{-3}$Gpc$^{-3}$ ($M>5\times 10^{13}\,h^{-1}M_{\odot}$).
{\bf Middle:} The same, but for for higher mass halo pairs merging within 
$\Delta t = 0.5$~Gyr of $z=0$ in the ($2.2\,h^{-1}$Gpc)$^3$ simulation.
The total number density of such mergers is
$1000\,h^{-3}$Gpc$^{-3}$ ($M>5\times 10^{13}\,h^{-1}M_{\odot}$),
$280\,h^{-3}$Gpc$^{-3}$ ($M>1\times 10^{14}\,h^{-1}M_{\odot}$),
$56\,h^{-3}$Gpc$^{-3}$ ($M>2\times 10^{14}\,h^{-1}M_{\odot}$).
{\bf Bottom:} The same as middle, but for a linking length of $b=0.20$.
Masses and virial radii have been scaled to that of $b=0.15$ for direct 
comparison.}
\label{fig:fmergemass}
\end{figure}

While it appears that cluster-mass halo pairs might provide a reasonable proxy
for mergers at $z \approx 1$, we next examine whether these results are robust
as a function of halo mass.
Figure~\ref{fig:fmergemass} (top) shows the pair merger fractions for halo
masses down to $M>5\times 10^{12}\,h^{-1}M_{\odot}$ at $z \approx 1$,
where the pair separation is shown as a multiple of the minimum mass cut
virial radius ($r_{500}$).
Figure~\ref{fig:fmergemass} (middle) shows the same but for higher halo masses
and a shorter time interval at $z \approx 0$.
While pairs of higher mass halos merge from larger physical pair separations, 
when scaled by the halo virial radius we find that the pair merger efficiency 
and completeness exhibit a nearly universal relation as a function of halo mass.
This implies that any of our results in \S\ref{sec:close} for a given mass 
can be approximately scaled to those of another mass.

This universal merger relation also implies that the total merger fraction
can be estimated by noticing that the following relation is satisfied at the
crossing point of the curves:
$(n_m(<r)/n_p(<r))/(n_m(<r)/n_{m,tot})=n_{m,tot}/n_p(<r)=1$. 
For example, from the middle panel of Fig.~\ref{fig:fmergemass} we find that
$n_{m\mbox{,tot}}=n_p(<r)$ when evaluated at a scaled pair separation of
$5r_{\mbox{vir}}$. One can thus estimate the merger rate by counting
pairs interior to $5r_{\mbox{vir}}$.
This is a potentially powerful result since it is approximately invariant for 
all mass-scales.\footnote{We thank the referee for pointing out this
universal behavior.}
However, this radius is a function of the timescale considered, and is likely to
depend on the cosmological parameters.
Furthermore, although their number densities agree, the pair population within
$5r_{\mbox{vir}}$ does not directly corespond to the total merger population. 
This makes it impossible to use the pair population to directly study any other
properties of mergers (e.g., clustering).
Redshift space distortions will significantly undermine the utility of these 
results~(see\S\ref{sec:redspace}).

The only strong deviation from this nearly universal mass relation occurs 
for merger completeness of $5\times 10^{12}\,h^{-1}M_{\odot}$ halos (top), 
which results in an intersection of merger efficiency and completeness at a 
lower fraction (10--15\% decrease) for lower-mass halos, a trend which we find 
does not depend strongly on redshift or merger timescale.
The use of pairs as proxies for merger rates becomes increasingly unreliable 
as we approach massive galaxy-size halos, an important result for massive 
\textit{galaxy} mergers at high-redshift, where galaxies are found primarily 
in distinct host halos.

These results augment those of \citet{Ber06}, who found that the evolution of
close galaxy pairs cannot be used to measure the host halo merger rate.
Specifically, they found that the host halo merger rate evolves as $(1+z)^3$,
while the number of close galaxy pairs evolves little with redshift.
This arises because, at low redshift, the merger rate of host halos is low, but
there are multiple galaxy pairs merging \textit{within} massive host halos.
At high redshift, the host halo merger rate is high, but the number density of
halos massive enough to host more than one bright galaxy is low.
Our results imply that even when considering major mergers of massive galaxies 
at high redshift, where each halo hosts a single bright galaxy, 
the close pair population will not reliably trace the merger population.

We have also compared two different linking lengths, to explore the dependence 
of our results on the FoF process.
In Fig.~\ref{fig:fmergemass} (bottom), we show the results using a linking 
length of $b=0.20$.
Since changing the linking length changes the mass of a given halo, we have 
scaled the mass threshold for $b=0.20$ to match the number densities of the
$b=0.15$ sample, thereby probing the same halo population. 
We have also scaled the halo separations by $r_{\mbox{vir}}=r_{500}$.
In the case of $b=0.20$, the intersection of the efficiency and completeness 
curves is shifted outward by $\sim r_{\mbox{vir}}$, which is not surprising 
since a halo found using $b=0.20$ is expected to be $\sim 30\%$ larger than 
one found using $b=0.15$.
From the figure we see that changing the linking length results in a change of 
only a few percent in the merger fraction at the intersection of the efficiency
and completeness curves. 
Thus changing the linking length does not qualitatively alter any of our 
results.
The weak dependence on the linking length suggests that for these mergers,
the FoF procedure does not give rise to significant artificial bridging of 
nearby halos, and our results would not differ significantly from a similar 
analysis performed using a spherical overdensity halo finder~\citep{lc94}.

\subsection{Merger Timescale Dependence of Pair Mergers}

\begin{figure}
\begin{center}
\resizebox{3.1in}{!}{\includegraphics{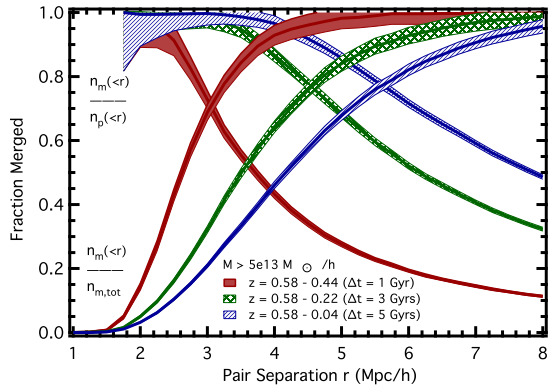}}
\end{center}
\caption{The fraction of pairs separated by less than $r$ that merged
(falling curves), indicating efficiency, and the ratio of identified
mergers to all mergers above the mass cut (rising curves), indicating
completeness, 
for various merger timescales from $z=0.58$.
The total number density of such mergers is
$950\,h^{-3}$Gpc$^{-3}$ ($z=0.58-0.44$),
$2755\,h^{-3}$Gpc$^{-3}$ ($z=0.58-0.22$),
$4282\,h^{-3}$Gpc$^{-3}$ ($z=0.58-0.04$).}
\label{fig:fmergezishalf}
\end{figure}

Pair merging is also strongly dependent on the choice of merger timescale.
Figure~\ref{fig:fmergezishalf} shows that the merger efficiency (falling curves)
increases with timescale, i.e.~pairs are more likely to merge when one considers
longer time intervals.
When considering mergers across $5$~Gyr, a significant fraction (80\%) of pairs
with separations $< 5\,h^{-1}$Mpc will merge.
However, the merger completeness within a given separation (rising curves)
decreases with timescale, so one must consider pairs at larger separation to
capture all mergers.
Interestingly, the trade off between efficiency and completeness is only weakly
dependent on the timescale, and we find similar results even for massive halo
pairs at $z = 1$ merging to $z = 0$ ($\Delta t = 7.6$~Gyrs).
No matter which merger timescale are considered, the utility of pairs
as proxies for mergers remains limited.

In all redshift, mass, and temporal regimes we consider, close halo pairs do
not provide a robust predictor of overall merger rates.
We conclude that, while pairs at small separations can reliably predict
mergers, these constitute a small fraction of the total merger population.
While our results could in principle be used to calibrate the merger fraction 
as a function of halo separation, close pairs do not provide a 
\textit{self-consistent} probe of the merger population independent of our 
theoretical predictions (which are contingent upon having simluated the correct 
cosmology).
\textit{Thus, even in the best-case scenario of complete knowledge of mass and 
position, measurements of galaxy cluster pairs cannot be used to measure 
cluster merger rates.
The same is true for any any class of object that singly occupies a halo,
particularly in using field galaxy pairs at high redshift to probe galaxy 
mergers.}

\section{Scatter in Mass, Redshift Space Distortions, and Redshift Errors}
\label{sec:redspace}

The situation becomes more complicated when we consider redshift space
distortions, redshift space errors, and scatter in the estimated mass of the
halos.
To identify the impact of imprecise determination of dark matter halo masses, 
we have recomputed several of the statistics of \S\ref{sec:close} after 
introducing an RMS scatter of $0.2 M_{\rm cut}$, where $M_{\rm cut}$ is the 
threshold for detecting the pairs in the mock observation. 
The scatter in the mass causes some halos to fall out of the sample and others
to enter it, resulting in a different population of halos near the threshold.
Because the mass function is steep, many more low-mass halos enter the sample
than high-mass halos leave it.  
Thus there are both more pairs to consider as merger candidates and more actual
mergers between members of the observed sample.
For example, if $M_{\rm cut}=5 \times 10^{13}\,h^{-1}M_{\odot}$, the number of
pairs increases more than the number of mergers, resulting in a decrease in
$n_m/n_p$ of $\sim5$\% for pairs separated by $2<r<6\,h^{-1}$Mpc over
$0.58<z<0.97$ ($\Delta t=2$~Gyrs).  
The effect is similar for other thresholds, redshifts, and intervals.
Scatter in the mass has a more pronounced impact at high values of 
$M_{\rm cut}$ where the mass function is steeper.  
The trend with redshift is derivative of the evolution of the mass function: at
a fixed value of $M_{\rm cut}$ there is more sensitivity to scatter at higher
redshifts where the mass function steepens.  
There is not a significant trend with merger timescale $\Delta t$.  
The overall impact of imprecise mass measurements is to degrade the range in
separation over which pairs might be considered useful proxies for mergers, 
but it is not a significant effect.

\begin{figure}
\begin{center}
\resizebox{3.1in}{!}{\includegraphics{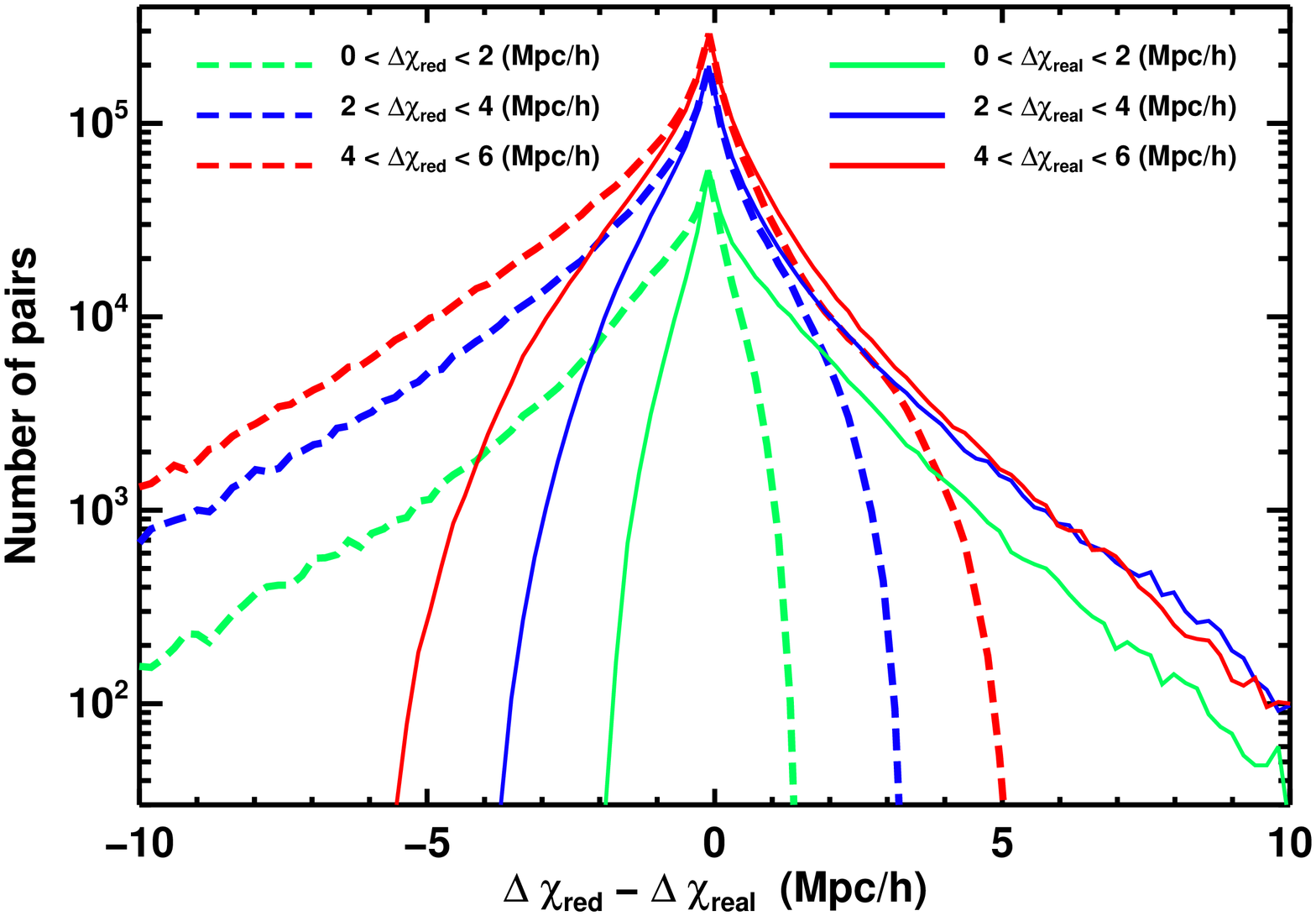}}
\end{center}
\caption{A histogram of the change in apparent separation because of 
redshift space distortions from halo peculiar velocities.  
Dashed lines indicate close pairs identified in redshift space (as in
observation) and solid lines indicate pairs identified in real space. 
The region $<0$ of the dashed lines indicate pairs that are artificially
close in redshift space.
The region $>0$ of the solid lines indicate pairs that are artificially
separated in redshift space.}
\label{fig:redbad}
\end{figure}

In contrast, redshift space distortions from doppler shift due to the halo
peculiar velocities $v_p$ are catastrophic for the pair-merger hypothesis.
When pairs are identified in redshift space, virtually none of them merge. 
In redshift space, the line of sight component is shifted relative to its
real position by an amount
\begin{equation}
\Delta\chi_{||}=\int_z^{z+v_p/c} \frac{c\, dz}{H(z)}.
\end{equation}
For typical halo velocities in our simulation, this shift amounts to a few
$h^{-1}$Mpc.
Figure~\ref{fig:redbad} illustrates that this distortion makes many close pairs
appear to be highly separated and pairs that are highly separated to appear
close. 
At redshift space separations less than $5 h^{-1}$Mpc, fewer than 5\% of merger
candidates actually merge in $\Delta t=0.6$ Gyr at $z=0$.
Even at $z=0.58$ where redshift space distortions are less severe, only 
$\sim$10\% of candidates merge over $\Delta t=1$Gyr.   
The situation improves modestly for longer time intervals, with $\sim$60\% of
pairs merging over $\Delta t=5.6$Gyr, but this falls far short of the fraction
in real space.
  
Figure~\ref{fig:redbad} shows a histogram of the difference between a pair's
separation in redshift space (dashed line), as would be done in an observation, 
and its true separation in configuration space (solid lines) at $z=0.04$.  
Merger candidates identified in real and redshift space are widely disjoint 
sets.  
The three sets of curves (colors) are bins of separation, as annotated. 
For the population of merger candidates identified in redshift space (dashed
lines), the region greater than zero shows pairs for which the real separation
was smaller than the measured separation, while the negative region shows pairs
that appeared to be close in redshift space but were actually separated by
several $h^{-1}$Mpc.
The former set might be likely to merge, but the latter set is highly unlikely
to do so.
For the actual population of close pairs (solid lines), the region greater than
zero shows pairs whose redshift space separations were greater than their real
separations, and which were therefore likely removed from the sample of merger
candidates.
The reason for the asymmetry between the pair population that left the sample
(solid, right side) and the population that entered the sample and replaced it
(dashed, left side) is that pairs are most often found in overdense regions,
toward which many other halos are streaming.
The impact is that the pair population identified in redshift space is larger
and has a broad distribution of physical separations; virtually none of the 
``merger candidates'' we identified actually merge.
This result is robust for all redshifts, merger intervals, and halo masses.

To a degree, this result also applies to the relationship between galaxy pairs
and galaxy mergers.
Observing galaxy pairs to study the merger rates of singly occupied host halos
at high redshift will be similarly affected by redshift space distortions, as
will using galaxy pairs to deduce the merger rates of multiply occupied host
halos at lower redshift.
However, the impact of redshift space distortions on using galaxy pairs to
deduce the galaxy-galaxy merger rates within a single host halo may be less 
severe \citep{Ber06}.  
Inside a halo, galaxy mergers are predominately between a satellite galaxy and
the central galaxy, and merger dynamics inside a host halo are influenced by 
dynamical friction, especially in group-mass halos and when the galaxy masses
are comparable.
Pair counts, even in redshift space, provide a rough proxy to the halo mass and
will therefore correlate well with the merger rate within the halo.

The impact of redshift space errors is similar to that of redshift space
distortions, except that, unlike with the velocities, the errors are not
correlated with the high density regions.
Redshift space errors that result in more than a few $h^{-1}$Mpc shift in
the apparent position have a similar impact to redshift space distortions.  

\section{The Merger Kernel and the Density Dependence of Mergers} \label{sec:kernel}

Since spatial information is a weak probe of halo merger statistics, 
we now turn to statistically describing the merger rate via the halo mass 
function.
For any population of objects that are built up by binary mergers of smaller 
constituents, the rate at which the number of objects of a given mass changes 
can be described by the Smoluchowski coagulation equation, which has been 
applied to the evolution of dark matter halos \citep{SilWhi78, CavColMen92, 
ShePit97, BenKamHas05}.
The abundance of halos at a given mass is increased through the creation of 
such halos by mergers of smaller halos, and decreased as these halos merge 
into more massive ones.
The rate of change of the number density of halos of a given mass can then be 
determined by knowing the number density of halos at all masses, i.e. the 
mass function, and the proper merger kernel to relate the mass function to a 
merger rate.
Historically, it has been difficult to study the coagulation of cluster-mass
halos through simulation because these events are rare.
However, with our large simulation volume the merger kernel can be computed in 
a statistically significant way. 

We define the merger kernel $Q(m_1,m_2,z)$ as in \cite{FurKam05} (hereafter
F\&K), via the relation
\begin{equation}
Q(m_1,m_2,z)=\frac{n_m(m_1,m_2,z)}{n(m_1,z)\,n(m_2,z)\,\Delta t},
\end{equation} 
where $n_m(m_1,m_2,z)$ is the number density of mergers between parents of mass
$m_1$ and $m_2$ in a time $\Delta t$, and $n(m,z)$ is the halo mass function. 
The quantity $Q$ can be interpreted as the efficiency of merging for a pair of
objects, such that the rate of mergers is a product of this efficiency with the
densities of available parents.  
Note that in previous contexts the term efficiency refers to the ability of the
close-pair technique to reliably identify merger candidates, whereas in this
context the efficiency is the coefficient of the number densities when
quantifying the merger rate. 
The time interval $\Delta t$ should be sufficiently short that there is no
significant evolution of the halo mass function. 
Satisfying this requirement drives down the number of mergers in a fixed 
volume, making statistically significant measurements difficult.
However, with our large simulation volume we measure $Q(m_1,m_2,z=0)$ on the
interval $0<z<0.04$ ($\Delta t = 0.6$~Gyr) by classifying the two progenitors
that contributed the most mass to each halo at $z=0$ into ten logarithmically
spaced mass bins in the range $10^{13} < h^{-1}M_{\odot} < 10^{15}$.  
The results are shown in Fig.~\ref{fig:merkernel}

We find that $Q$ follows the simple functional form 
\begin{equation}\label{eqn:qform}
Q(m_1,m_2,z=0)=A \, \left[\frac{m_1+m_2}{h^{-1}M_\odot}\right]^B \, .
\end{equation}
The best fit values for $A$ and $B$, found using a linear least-squares fit 
to the data, are presented in Table~\ref{tab:merkernel} (bottom row).
This functional form satisfies the formal requirement that the merger kernel be
symmetric in its two arguments.
For comparison, \citet{BenKamHas05} analytically found that 
$Q \propto (m_1+m_2)$ when $P(k) \propto k^n$ with $n=0$, which is approximately
true in the trans-linear regime of cluster-mass halos.

\begin{table}
\begin{center}
\begin{tabular}{|c|cccc|}
\hline
$\bar{\delta_i}$ & $V$ ($h^{-1}$Gpc)$^3$ & $A$ (\,$h^{-1}$kpc)$^3$/Gyr  
& $B$ & $\chi_{\rm red}^2$ \\
\hline
$-0.56$  & $.47$ &  $0.121 \pm 1.99 \times 10^{-3}$ & $0.88$ & $1.00$ \\
$-0.21$  & $.33$ &  $0.086 \pm 1.30 \times 10^{-3}$ & $0.88$ & $1.55$ \\
$ 0.20$  & $.27$ &  $0.072 \pm 9.98 \times 10^{-4}$ & $0.88$ & $1.46$ \\
$ 1.33$  & $.21$ &  $0.060 \pm 7.10 \times 10^{-4}$ & $0.88$ & $1.41$ \\
\hline
$0.0$ & $1.28$ & $.088 \pm 2.23 \times 10^{-2}$ & $0.88\pm 0.008$ & $3.09$ \\
\hline
\end{tabular}
\end{center}
\caption{Best fit amplitude, $A$, and exponent, $B$, for the merger kernel 
$Q = A(m_1+m_2)^B$, where $(m_1+m_2)$ is the sum of the 2 progenitor masses. 
The first four rows are the results for each of the 4 density subdivisions,
while the bottom row is the best fit for the entire simulation.
The average density and total volume of each subdivision is listed at left,
and the reduced $\chi^2$ for each subdivision is listed at right.}
\label{tab:merkernel}
\end{table}

\subsection{Density Dependence of the Merger Kernel}

Because the densities of the parent populations have been divided out, it is
conceivable that the merger kernel $Q$ is independent of the large-scale 
density field.
Significant dependence on density would indicate that environmental effects
other than the progenitor densities, such as halo velocity distributions, halo
impact parameters, and tidal fields, are at play in determining the merger rate.
To investigate the dependence of $Q$ on the large-scale density field we
construct a coarse ($64^3$) density grid in our simulation and assign the dark
matter particles to the nearest grid point, effectively smoothing the field on 
a scale of $\sim17\,h^{-1}$Mpc.  
Each halo at $z=0$ is also assigned to the coarse grid, and these halos are
sorted in order of their large scale density environment and then divided into
quartiles of density, with each quartile containing approximately the same
number of halos.
Using the coarse grid, we compute the total volume and overdensity of each
quartile in the simulation, and these appear in the first two columns of Table
\ref{tab:merkernel}.  
The mean overdensity of a quartile is defined as
\begin{equation}
\left<\delta_i\right>=\frac{\left<\rho \right>_i-\left<\rho \right>}
{\left<\rho \right>} \hspace{1cm} i=1,2,3,4
\end{equation}
where the angular brackets denote an average of the $\sim$($17\,h^{-1}$Mpc)$^3$
cells of the coarse grid.
The merger kernel $Q$ is fit for each quartile to the form of
Eq.~\ref{eqn:qform}, using the value of $B$ determined from the entire
simulation.
The results are summarized in Table \ref{tab:merkernel}.
The best fit values of $A$ differ by several sigma, and the improvement in the
reduced $\chi_{\rm red}^2$ statistic indicates that each of the individual
density fits is a much better fit that a fit to the entire volume.
This indicates that there is a clear trend in $Q$ with the large-scale density
field: as the density increases the efficiency of merging for a given system
mass decreases.
\textit{Thus, the environmental effects in dense environments are hindering
the merging process in comparison to less dense environments.}

\begin{figure}
\begin{center}
\resizebox{3.1in}{!}{\includegraphics{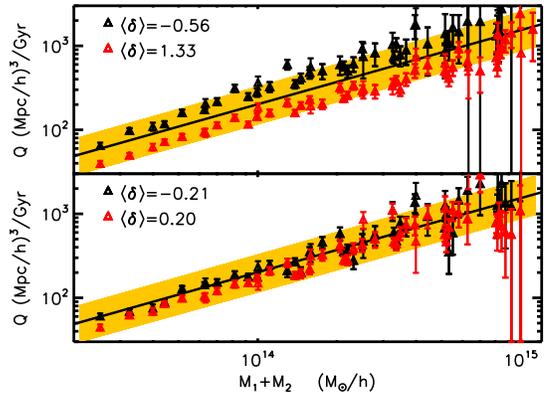}}
\end{center}
\caption{Merger kernel $Q$ from the simulation for 10 logarithmic mass bins
in progenitor masses, as a function of the sum of the progenitor masses.  
The solid line depicts the best fit of Eq.~\ref{eqn:qform} to the whole
simulation volume.  
The shaded region corresponds to the formal $1\sigma$ errors on the fit
values of $A$ and $B$.  
The top panel shows $Q$ computed in the highest and lowest density quartiles 
and the bottom panel shows the quartiles closer to the mean density. 
The highest and lowest density regions differ in $Q$ by several sigma, and the
density dependence of $Q$ is one of the main factors driving the dispersion in
the fit to the whole box (shaded region).} 
\label{fig:merkernel}
\end{figure}

Figure~\ref{fig:merkernel} shows $Q$ in each of the four density environments
plotted individually. 
The upper panel shows $Q$ in the highest and lowest density quartiles, while the
lower panel shows the results from the inner two quartiles.  
The solid line is the best fit model for the entire box, which is only a good 
approximation for environments close to the mean density.
The yellow shaded region results from allowing $A$ and $B$ to deviate by their
$1\sigma$ values.
Figure~\ref{fig:merkernel} demonstrates clearly that a large component of the
dispersion in the best fit to the entire simulation originates in the density
dependence of $Q$.

Figure~\ref{fig:merkernel} clearly indicates that merging is less efficient in
high density environments than in low density environments, but we note that
with regard to our fit there is an ambiguity as to whether this implies a lower
normalization $A$ or a shallower power law $B$.  
Given the rarity of these events at the high mass end and the limited dynamic
range that is consequently driving the fit, there is large covariance between
the fit values of $A$ and $B$.
We have performed the 2-parameter fit to Eq. \ref{eqn:qform} for each 
of the density quartiles individually and find that indeed $A$ increases
slightly with density (in contrast to the 1-parameter fit) while $B$ decreases.
Unfortunately, the formal errors in the parameters (from inversion of the
covariance matrix) become so large that the four regions are statistically
indistinguishable.
However, the individual fits hint at an interesting possibility: if $B$
decreases and $A$ increases with density, then the merger efficiencies
cross over as a function of system mass.
Specifically, for lower total system masses ($m_1+m_2$), merging becomes more
efficient in high density environments.
In our simulation, all four $Q$ curves cross over in the mass range
$10^{10}\,h^{-1}M_{\odot}<(m_1+m_2)<10^{12}\,h^{-1}M_{\odot}$, suggesting that
galaxy scale mergers may be more efficient in denser environments.
While this evidence is extrapolated from higher mass, the trend is potentially
worth further investigation given that other halo properties, e.g.~clustering 
as a function of concentration/formation time, reverse trend from $M > M_*$ to 
$M < M_*$\citep{GaoSprWhi05,Wec06}\footnote{
$M_*(z)$ is the mass at which $\sigma(M)$, the variance of the linear
power spectrum smoothed on scale $M$, equals the threshold for linear density
collapse $\delta_c(z)$; $M_* \approx 8\times10^{12}\,h^{-1}M_\odot$ at $z=0$.}.

We note that the merger kernel $Q$ as computed in this section is not a direct
cosmological observable.
There is an observational counterpart to the merger kernel: the pair kernel,
$Q_p$, computed using the number density of pairs with separations 
$r<r_{\Delta t}$.  
The threshold separation is calibrated from simulations to any desired 
tolerance for completeness or contamination, but the result is insensitive to
$r_{\Delta t}$.
In all cases, there are many more pairs than actual mergers, the amplitude $A$
of $Q_p$ is larger, and the power law $B$ is shallower, both by several
$\sigma$.  
It  infeasible to use pairs at any mass scale as a proxy to compute merger
rates.

\subsection{Density Dependence of Close Pair Mergers}

\begin{figure}
\begin{center}
\resizebox{3.1in}{!}{\includegraphics{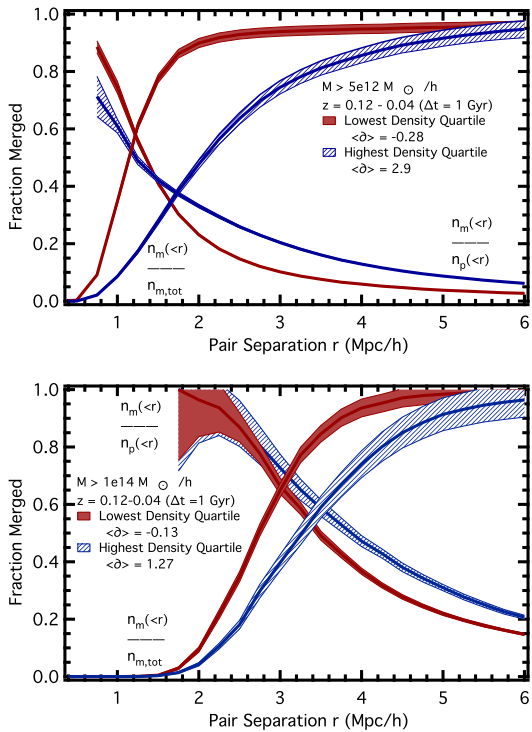}}
\end{center}
\caption{The fraction of the total number of pairs separated by less than $r$
that merged (falling curves), indicating efficiency, and the fraction of all
mergers in the simulation within $\Delta t = 1$~Gyr (rising curves), indicating
completeness, for the highest and lowest density quartiles.
Pairs are found at $z=0.12$ for $M>5\times 10^{12}\,h^{-1}M_{\odot}$ (top)
and $M>10^{14}\,h^{-1}M_{\odot}$ (bottom).}
\label{fig:fmerge_density}
\end{figure}

The trend in the merger kernel is driven by the fact that the number of mergers
grows more slowly with density than the parent mass functions.   
By studying changes in the relationship between close pairs and mergers in 
regions of differing local density, we can demonstrate that this is not 
universally true for all pair separations.  
We proceed in an analogous manner, smoothing over $\sim17\,h^{-1}$Mpc and
defining four density quartiles, each containing the same number of mergers.
We consider pairs merging between $z=0.12$ and $z=0.04$ ($\Delta t = 1$~Gyr), 
though our results are insensitive to the time interval and redshift. 
Fig.~\ref{fig:fmerge_density} shows the results for pairs of halos above
$5\times 10^{12}\,h^{-1}M_{\odot}$ (top) and $10^{14}\,h^{-1}M_{\odot}$
(bottom).
The relationship between close pairs and mergers changes with the large-scale
density.
Pairs within a given separation are a less complete sample of merger candidates
in overdense environments than in underdense regions (rising curves, top and
bottom).
Objects in overdense regions have higher velocities, which allow mergers to 
come from a broader range of progenitor separations and can extend the infall 
time of closer pairs by generating larger angular momenta.

Figure~\ref{fig:fmerge_density} also indicates that, at large separations, 
pairs are more efficient predictors of mergers in overdense regions than in 
underdense regions (falling curves) because higher velocities in overdense 
regions allow pairs from larger separations to merge.
However, at small separations ($\lesssim 2\,h^{-1}$Mpc), the relationship
between pairs and mergers depends on the pair masses.  
Here, there are no high-mass ($\gtrsim10^{14}\,h^{-1}M_{\odot}$) halos because
of halo exclusion, and pairs are more efficient predictors of mergers in {\it
underdense} regions (top).
This is likely because low-mass progenitors in overdense regions have large
velocities and thus have longer infall times,  e.g.~from tidal fields, resulting
in large angular momenta that inhibit merging.
In contrast, the smaller velocities in underdense regions enhance the
probability that a very close pair will merge from simple gravitational
attraction.
High-mass halos have much stronger gravitational attractions and are thus less
affected by the large-scale density field.
Indeed, at small separations, massive halos are equally likely to merge in 
over- and underdense environments.

Finally, we note that the two methods we have presented to examine the 
density dependence of merging probe mergers in different ways.
The merger kernel integrates over merger pair separations, thus considering
mergers as a function of progenitor pair mass sum.
Alternately, the close pair merger method selects only close halo pairs that 
are above a given mass cut, thus integrating over mass dependence and 
considering mergers as a function of pair separation.
The merger kernel indicates that for all child halo masses in our simulation
merging is more efficient in underdense environments.
This is corroborated by Fig.~\ref{fig:fmerge_density} (top), which shows that
for all halos in our simulation, pair merging is more likely in underdense
environments for pairs within $1.5\,h^{-1}$Mpc separation.
The accompanying merger completeness (rising) curves show that pairs within 
this separation constitute the majority of pair mergers in the given timescale.
Figure~\ref{fig:fmerge_density} (bottom) shows that high-mass pairs are 
more likely to merge in overdense regions at all pair separations.
However, since mergers between two halos both with 
$M>10^{14}\,h^{-1}M_{\odot}$ are extremely rare (as opposed to mergers between
a high and low mass object), these contribute less to the merger kernel at a 
given child mass.

\section{The Clustering of Close Pairs and Merger Bias}
\label{sec:xipairs}

For two halos to merge, they must have been located in close physical proximity
at an earlier time.
Since closely spaced halos are more likely to be found in overdense regions, 
recently merged halos may exhibit enhanced spatial clustering.
Moreover, if recently merged halos cluster differently from the general
population (``merger bias'') and this is unaccounted for, conclusions drawn
about halos on the basis of their clustering could be compromised.
For example, the use of cluster self-calibration to infer cluster masses
(and in turn cosmological parameters) requires a precise knowledge of the 
clustering of clusters as a function of mass \citep{MajMoh03}.
A number of earlier studies have looked for such merger bias
\citep{Got02,Per03,ScaTha03,Wet07}, with mixed results.

Recently, \citet{FurKam05} developed an analytic model to predict the
merger bias.
Assuming that mergers correspond to closely spaced objects at an earlier time,
they compare the clustering of close pairs to that of single objects, thus
computing the pair bias as a proxy for merger bias.
On scales much larger than the pair separations, they found an enhancement of
clustering for pairs of mass $M>M_*$, implying that recently merged high-mass
halos should exhibit a clustering bias.
We now use F\&K's framework to determine if the clustering of close halo pairs
of mass $M \gg M_*$ provides an accurate proxy for the clustering of mergers.

Using simulations, \citet{Wet07} found 
the most prominent merger bias ($\sim20$\%) for major mergers ($M_2/M_1>0.3$) 
of high-mass halos ($M>5\times10^{13}\,h^{-1}M_{\odot}$) at $z=0$ over 
$\Delta t=0.6$~Gyr. 
We consider a similar mass and temporal regime.
To improve our statistics we additionally use a box of eight times the volume
previously used (see \S\ref{sec:sim}), which allows us to probe merger effects
of more massive halos.
We use a shorter timescale of $\Delta t=0.5$~Gyr at $z=0$ to define our merger
interval to preserve the signal in the comparison population.
Contrary to expectation, the merger bias \textit{does not} increase with mass;
it remains a 10--20\% enhancement with similar statistical significance up to
halos of mass $M>4\times10^{14}\,h^{-1}M_{\odot}$.

In computing the pair bias, F\&K define ``close'' halo pairs by demanding that
the probability of finding three or more halos in a sphere of a given radius is
small compared to that of finding two.
This is approximately equivalent to the restriction that the probability of
finding two or more neighbors, $P(\geq2)$, within a given separation from a 
halo is small compared to that of finding one, $P(1)$, which is what we will 
measure.
For halos $M>5\times10^{13}\,h^{-1}M_\odot$ at $z \approx 0$, we find that
$P(\geq2)/P(1) \approx 0.1$ for a (comoving) sphere radius of $4\,h^{-1}$Mpc.
This separation restriction yields $\sim6,800$ pairs per $(h^{-1}$Gpc$)^{3}$,
out of $\sim77,000$ halos per $(h^{-1}$Gpc$)^{3}$.
While this is a sufficient number density of pairs for a robust correlation
function measurement in our ($2.2\,h^{-1}$Gpc)$^3$ simulation volume,
\S\ref{sec:close} demonstrated that close pairs do not reliably predict the
merger population.

The analytical model of F\&K predicts a significant merger bias in its
application to the clustering of massive halos ($M \gg M_*$).
For such objects, it predicts a correlation function of pairs $X_p(r)$ in terms
of $\xi_h(r)$, the correlation function of individual halos:
\begin{equation} \label{fk} 
X_p(r)=[1+\xi_h(r)]^{4}-1.
\end{equation} 
This is computed on scales where the underlying matter fluctuations are linear,
which at $z=0$ corresponds to distances greater than $\sim10\,h^{-1}$Mpc.  
To compute the pair bias F\&K use
\begin{equation}
b_p^{2} \equiv \frac{X_p(r)}{\xi_h(r)}.
\end{equation} 
For a given halo mass, this can result in anomalously high pair biases.  
This is because the halo correlation function is implicitly a function of the
halo mass; more massive halos are more highly clustered.   
When computing the merger bias (or its proxy, the pair bias) the comparison
population should be an ensemble of halos of the same characteristic mass as 
the child halo, not the parent halos.  
Otherwise, the merger bias becomes entangled with the effect that larger halos
are more clustered. 
To adjust for this in the analytic model we note that 
$\xi_h(r,2M) \approx 1.1 \xi_h(r,M)$ near $M \sim10^{14} M_\odot$.
On scales $>10\,h^{-1}$ Mpc, $\xi(r)$ is less than $1$,  so this leads to a 
pair bias of $b_p^{2} \approx 3.6$.
This pair bias is still significantly larger than the 10--20\% merger bias seen
in simulation for halos up to $M>4\times10^{14}\,h^{-1}M_\odot$. 

As an alternative to an analytic approach, we next measure the clustering of
close pairs in simulations to discover whether pair bias can predict merger
bias. We consider pairs $0.5$~Gyr prior to $z=0$.
To assign a unique position to a pair of neighboring halos, we impose a
``couples'' restriction, namely that each member of a pair is the other's
nearest neighbor.
This restriction remains robust for two body mergers, a good approximation for
our time interval.
Within the pair separations we consider ($<2\,h^{-1}$Mpc) couples constitute
essentially all pairs.
We select couples of halos above $5\times 10^{13}\,h^{-1}M_{\odot}$ at 
$z=0.04$ and use their geometric centers to evaluate the couple correlation
function.
We use couples with separations less than $1.6\,h^{-1}$Mpc, 80\% of which
correspond to mergers in our time interval (Fig.~\ref{fig:fmergemass}).
To limit the effect of mass-dependence on the halo correlation function, we
compare this correlation function with that obtained from single halos above
$10^{14}\,h^{-1}M_{\odot}$ at $z=0$.
While this ignores the effects of mass scatter in mergers, such effects remain
small for this short timescale.

\begin{figure}
\begin{center}
\resizebox{3.1in}{!}{\includegraphics{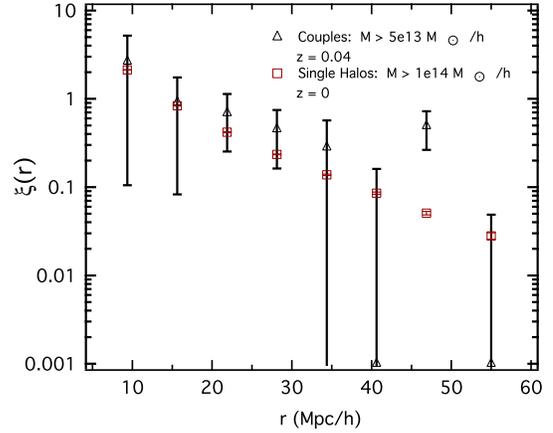}}
\end{center}
\caption{Correlation function of couples (triangles) at $z=0.04$
compared with that of single halos (squares) at $z=0$.
Mass cuts are chosen so that the single halos are approximately the same
mass as the child products of the couples.
While enhanced signal is seen at $47h^{-1}$Mpc, this is mitigated by the
adjacent points which are both negative.}
\label{fig:xipairs}
\end{figure}

Figure~\ref{fig:xipairs} shows that with a ($2.2\,h^{-1}$Gpc)$^3$ simulation
volume, no statistically significant pair bias is found.
Such poor statistics arise because only a small fraction (15\%) of total 
mergers are represented by couples at such close separations.
Trying to increase statistics by considering couples at larger separations is
undermined by the fact that the fraction of couples that merge is a steeply
decreasing function of separation.
As already mentioned, one cannot consider longer merger timescales, as this
permits a larger fraction of the halo population to have undergone a major
merger, thereby washing out the signal in the comparison population.
We find similar results when looking at the pair bias for mass cuts from 
$M>10^{13}\,h^{-1}M_{\odot}$ to $M>2\times 10^{14}\,h^{-1}M_{\odot}$, 
and thus we conclude that \textit{pair bias cannot be used to reliably predict
merger bias observationally, in simulation, or through current analytic
treatment.}

\section{Conclusions}\label{sec:conclusions}

Cluster merger statistics may provide insight into the nature of hierarchical
structure formation and the mechanisms by which the largest coherent objects in
the universe form.
We use large-volume, high-resolution N-body simulations to investigate the
utility of close spatial pairs of galaxy clusters as proxies for cluster
mergers.
We characterize merger statistics through the merger kernel, and examine the
density dependence of merger efficiency.
We highlight our conclusions as follows:

\begin{itemize}
\item Close pairs of galaxy clusters at very small separations 
($< 1-2\,h^{-1}$Mpc) can be used to reliably predict mergers.  
However, since these constitute a small fraction of the total merger population,
close pairs are not a reliable proxy for cluster merger rates.
We quantify this by measuring the efficiency and completeness of merger
candidates identified via close pairs as a function of separation and find that
their intersection typically occurs at a low merger fraction ($0.5$--$0.6$).

\item We find that close pairs are even poorer proxies for mergers between 
massive galaxy-sized halos. 
This indicates that galaxy pairs will not provide a reliable proxy for galaxy 
merger rates at high redshift, where most galaxies reside in distinct halos.

\item We note that the failure of close pairs as proxies for mergers indicates
that determination of merger rates from spatial statistics, such as the
correlation function, cannot be trusted outside the physical size of a single
halo.

\item We examine the mass, redshift, and timescale dependence of pair mergers,
finding that the pair-merger hypothesis (that close pairs are proxies for 
mergers) at a given comoving separation is most accurate at high redshift, 
high mass, and over long merger timescales.  
In our best scenarios, the intersection of merger efficiency and completeness 
is at $~75$\%, i.e.~75\% of pairs within a given separation merge, and these 
constitute 75\% of all mergers.
We also find a nearly universal relation for pair merger efficiency and 
completeness for different mass halos. 
This relation begins to break down as we approach massive galaxy-size halos, 
and is compromised by redshift space distortions.

\item Redshift space distortions have a devastating impact on detecting close
galaxy cluster pairs in surveys; nearly all of the merger candidates identified
in redshift space do not merge.
Although an extrapolation, we expect these results to be robust for galaxy-size
halos at high redshift.

\item We present the first fit from simulation to the merger kernel---a means 
to describe halo merger rates via the halo mass function (coagulation).

\item The merger kernel exhibits dependence on local ($\sim17\,h^{-1}$Mpc)
density.
Specifically, halo merging in our high-mass regime is more efficient in
underdense regions.

\item Pairs at large separations ($\gtrsim 3\,h^{-1}$Mpc) are more likely to
merge in overdense regions.
For pairs at small separations, low-mass halos are more likely to merge in
underdense regions, while high-mass halos exhibit no environmental
dependence.

\item  We sought to use cluster pairs to measure the merger bias by using the
pair bias as a proxy for merger bias.
We extended the treatment of previous analytic work to include the fact that
mergers result in mass gain; when computing the bias of recently merged halos,
the comparison population should be a set of halos of the same mass as the
children, instead of the parents.
Close spatial pairs that reliably merge are too rare to produce a statistically
significant measure of merger bias, even in a ($2.2\,h^{-1}$Gpc)$^3$ simulation
volume.
\end{itemize}

In conclusion, we have shown that close spatial pairs of galaxy clusters are of 
limited value as a probe of overall cluster merger rates.
We have determined the merger kernel for halo coagulation for the first time 
from simulations, finding that a statistical description of halo mergers is of 
more promise.
Further work is needed to extend our parametrization of the merger kernel to 
lower masses and higher redshifts, and to explore whether the environmental 
dependence of the merger rate persists in these regimes.

\begin{acknowledgments}
We thank G. Jungman and R. Yan for enlightening conversations, and J. Cohn, S.
Furlanetto, M. Kamionkowski, and M. White for valuable comments on an early 
draft.
Computational resources were provided by the LANL open supercomputing initiative
and the Space Simulator Beowulf cluster.
A.S. and A.W. were supported by NASA.
\end{acknowledgments}

\end{document}